\documentclass{emulateapj}
\slugcomment{Accepted by {\it The Astrophysical Journal Letters}}
\def\la{\mathrel{\hbox{\rlap{\hbox{\lower4pt\hbox{$\sim$}}}\hbox{$<$}}}}
\def\ga{\mathrel{\hbox{\rlap{\hbox{\lower4pt\hbox{$\sim$}}}\hbox{$>$}}}}

\shortauthors{Post}
\shorttitle{SNR G299.2-2.9}

\begin{document}
\title{Asymmetry in the Observed Metal-Rich Ejecta of Galactic Type Ia Supernova Remnant G299.2-2.9}

\author{Seth Post\altaffilmark{1}, Sangwook Park\altaffilmark{1},
Carles Badenes\altaffilmark{2}, David N. Burrows\altaffilmark{3}, John P. Hughes\altaffilmark{4}, Jae-Joon Lee\altaffilmark{5}, Koji Mori\altaffilmark{6} Patrick O. Slane\altaffilmark{7}} 

\altaffiltext{1}{Box 19059, Department of Physics, University of Texas at Arlington,
Arlington, TX 76019; seth.post@mavs.uta.edu}
\altaffiltext{2}{Department of Physics and Astronomy and Pittsburgh Particle Physics, 
Astrophysics, and Cosmology Center (PITT-PACC), University of Pittsburgh, 3941 O’Hara 
Street, Pittsburgh, PA 15260, USA; badenes@pitt.edu}
\altaffiltext{3}{Department of Astronomy and Astrophysics, Pennsylvania State
University, 525 Davey Laboratory, University Park, PA 16802; burrows@astro.psu.edu}
\altaffiltext{4}{Department of Physics and Astronomy, Rutgers University,
136 Frelinghuysen Road, Piscataway, NJ 08854-8019; jph@physics.rutgers.edu}
\altaffiltext{5}{Korea Astronomy and Space Science Institute, Daejeon, 305-348, Korea}
\altaffiltext{6}{Department of Applied Physics, University of Miyazaki, 1-1 Gakuen 
Kibana-dai Nishi, Miyazaki, 889-2192, Japan; mori@astro.miyazaki-u.ac.jp}
\altaffiltext{7}{Harvard-Smithsonian Center for Astrophysics, 60 Garden Street,
Cambridge, MA 02138; slane@cfa.harvard.edu}

\begin{abstract}

We have performed a deep {\it Chandra} observation of Galactic Type Ia supernova remnant G299.2-2.9.  Here we report the initial results from our imaging and spectral analysis.  The observed abundance ratios of the central ejecta are in good agreement with those predicted by delayed-detonation Type Ia supernovae models.  We reveal inhomogeneous spatial and spectral structures of metal-rich ejecta in G299.2-2.9.  The Fe/Si abundance ratio in the northern part of the central ejecta region is higher than that in the southern part.  A significant continuous elongation of ejecta material extends out to the western outermost boundary of the remnant.  In this western elongation, both the Si and Fe are enriched with a similar abundance ratio to that in the southern part of the central ejecta region.  These structured distributions of metal-rich ejecta material suggest that this Type Ia supernova might have undergone a significantly asymmetric explosion and/or has been expanding into a structured medium.  

\end{abstract}


\section {\label {sec:intro} INTRODUCTION}

Type Ia supernovae (SNe) are the thermonuclear explosions of carbon-oxygen (CO) white dwarfs within close binary systems. When a CO white dwarf approaches the Chandrasekhar limit through mass accretion from a non-degenerate companion star ({\it single--degenerate} [SD] channel) it becomes unstable and a thermonuclear runaway occurs.  A thermonuclear explosion may also be produced by the merging of two white dwarfs ({\it double--degenerate} [DD] channel).  The detailed physics involved in these thermonuclear explosions and the nature of their progenitor systems are under debate, and several models exist to describe Type Ia SN explosion mechanisms (e.g., Maoz et al. 2014 for a recent review).  While most Type Ia SNe expand into uniform surroundings (e.g., Badenes et al. 2007), there are a growing number of observational and theoretical studies which indicate that some Type Ia SNe could occur in surroundings modified by stellar winds from the progenitors or binary companions (e.g., Hamuy et al. 2003, Hughes et al. 2007, Dilday et al. 2012, Maguire et al. 2013).  The specifics of the local environment may depend upon the nature of the progenitor system.   
	 
Studies of supernova nucleosynthetic products and the ambient structure of Type Ia supernova remnants (SNRs) can be helpful to reveal the detailed nature of their progenitor systems and explosion mechanisms.  Such nucleosynthesis studies can be effectively performed with spatially-resolved spectroscopy of ejecta-dominated Type Ia SNRs.  The Galactic SNR G299.2-2.9 is an excellent laboratory for such a study.  G299.2-2.9 was discovered by the {\it ROSAT} all sky survey \citep{buss95}.  A {\it Chandra} study found a central region enhanced in Fe, Si, and S, which indicated a metal-rich ejecta origin from a Type Ia SN \citep{park07}.  G299.2-2.9 is significantly evolved (age $\gtrsim$4500 yrs, Slane et al. 1996, Park et al. 2007), and thus the reverse shock has probably heated the bulk of metal-rich ejecta in this SNR.  G299.2-2.9 exhibits a complex X-ray morphology with multiple shell-like structures \citep{park07}.  Spectrally-soft faint diffuse emission features extend beyond the bright shells almost all around the SNR.  This complex structure suggests that G299.2-2.9 might have exploded in a non-uniform environment, possibly with a density gradient generally along the line of sight \citep{park07}.  It has also been speculated that the asymmetric morphology of the outer shell might have been caused by bi-polar outflows of stellar winds (as seen in planetary nebulae) from the progenitor or the companion \citep{tseb13}. Thus, G299.2-2.9 provides an excellent opportunity to study a Type Ia progenitor that exploded in a modified environment.

Based on our deep {\it Chandra} observation we have commenced a study of the ejecta and the circumstellar material of G299.2-2.9.  In this  {\it Letter} we report on the initial results from our ejecta study focusing on its spatial structure.  Our results from more extensive studies (including the outer shell structures) will be presented in our follow-up work. In Section 2 we describe our observations.  Our data analysis is presented in Section 3.  We present a discussion of our results in Section 4.

\section{\label{sec:obs} OBSERVATIONS}

We performed our observations of G299.2-2.9 with the Advanced CCD Imaging Spectrometer (ACIS; Garmire et al. 2003) on board {\it Chandra} between 2010 October 26 and 2010 November 13.  A total of nine ObsIDs were obtained using the ACIS-I array in Very Faint mode.  We performed data reduction with Chandra Interactive Analysis of Observations (CIAO) version 4.5 and CALDB version 4.5.1.    We did not find severe variability in the background light curve. We corrected the spatial and spectral degradation of the ACIS data caused by radiation damage, known as the charge transfer inefficiency \citep{town00}. We carried out standard data screening by status, grade, and energy selections. We removed ``flaring'' pixels and selected ASCA grades (02346).  The total effective exposure is $\sim$628 ks after the data reduction. The overall SNR spectrum is soft with few source photons above $E$ $\sim$ 3 keV. At low energies ($E$ $\lesssim$ 0.4 keV), the source flux is negligible because of the foreground absorption, and X-ray emission is dominated by the detector background. Thus, we extracted photons between 0.4 and 3.0 keV for each observation in our data analysis.

\section{\label{sec:results} DATA ANALYSIS}

Combining all ObsIDs, we detected $\sim$190 faint point-like sources within the ACIS-I field of view using the wavdetect script in CIAO.  We removed them before any further data analysis.  We present our X-ray 3-color image of G299.2-2.9 in Fig. 1a.  In this 3-color image we replaced the source regions identified by our point source detection with count values from the Poisson distribution of the area surrounding each region.  Based on these {\it Chandra} data, we reveal the detailed X-ray morphology of this SNR in its entirety\footnote{The entire G299.2-2.9 was also detected in archival {\it XMM-Newton} data (ObsIDs 0112890101 and 0112890201, with a total exposure of $\sim$28 ks), but these data have not been published.}. We also reveal the entire morphology of the central ejecta region.  The outermost angular extent of G299.2-2.9 is $\sim$13$'$ from east to west while that from north to south is $\sim$11$'$.  The angular extent of the bright inner shell from north to south is $\sim$8$'$ and from east to west is $\sim$11$'$.  The main part of the central ejecta region is roughly circular with an extent of $\sim$4.5$'$ in diameter.  This central ejecta region is overall dominated by emission in the 0.72-1.4 (green) and 1.4-3.0 keV (blue) bands (Fig. 1a) whereas the 1.4-3.0 keV band emission is emphasized in the southern half of the central ejecta region.  This spectrally-hard emission extends continuously from the central ejecta region to the western outermost boundary of the SNR (Fig. 1a).  This significantly elongated (to the west) emission feature was not detected in the previous {\it Chandra} study because the ACIS-S3 (with a smaller field of view) was used there, resulting in incomplete coverage of the SNR \citep{park07}.  

We constructed line {\it equivalent width} (EW) images for the prominent emission lines (Fe-L and Si-K, Figs. 2a \& 2b) following the methods described in literature (e.g., Hwang et al. 2000, Park et al. 2002).  These EW images help us map regions where line emission is enhanced across the SNR.  We note that the use of EW images is only a qualitative guide to identify line-enhanced/suppressed areas for an efficient regional spectral analysis.  The Si EW image is enhanced (relative to Fe) in the southern half of the central ejecta region.  This enhancement continuously extends to the outermost boundary of the SNR in the west.  The marginally-enhanced Si EWs in the eastern parts of the SNR do not represent Si overabundance based upon our spectral model fits for those regions (not shown in this work). The moderate Si EWs in the eastern boundary of the SNR correspond to the swept-up circumstellar material (CSM) and/or ISM shell (with sub-solar abundances) identified by Park et al. (2007).  This dense CSM/ISM material surrounds (in projection) the central metal-rich ejecta (as shown by Park et al. 2007), except to the West.  The Fe EW is centrally enhanced and also appears to extend to the western outermost boundary of the SNR.  We note that due to potential Ne contamination from the shell the Fe-L EW image may not accurately map the entire Fe emission, particularly along the outer shell regions.  However, the Fe EW map should be reliable in the central ejecta region as the Ne line is negligible there.  Similar distributions of Si and Fe line emission are evident in the Fe/(O+Mg) and Si/(O+Mg) line ratio maps (Figs. 2c \& 2d).  In contrast to Fe and Si lines which primarily trace the metal-rich ejecta, the O and Mg lines originate mostly from the shocked ambient medium in G299.2-2.9 (see Park et al. (2007) and below for relevant discussion).  These line ratio maps clearly demonstrate that Fe and Si line emission is enhanced, compared to O+Mg, in the central ejecta region, and extends primarily to the western SNR boundary (Figs. 2c \& 2d).  These Fe and Si line enhancements are anti-correlated with the broadband intensity contours which are dominated by emission from the bright swept-up shell.  This anti-correlation is consistent with the ejecta origin for the enhanced Fe and Si line emission.      

We have examined X-ray spectra from many regions throughout this SNR, and have selected three characteristic regions (North, South, and West, Fig. 1a) to highlight differences seen in the ejecta structure as suggested by our EW images and line ratio maps. Based on our spectral analysis of several regions in the bright shells and outermost faint emission regions, we selected the Shell region (Fig. 1a) to represent the shocked ambient medium.  We extracted spectra from these individual regions from each ObsID and then merged them together using specextract in CIAO.  Each merged spectrum contains $\sim$4000--8000 counts.  We binned each regional spectra to contain a minimum of 20 photons per channel.  We fit each spectrum with a non-equilibrium ionization (NEI) plane-parallel shock model (vpshock with NEI version 2.0 in Xspec, Borkowski et al. 2001)  based on the ATOMDB (for a description of ATOMDB, see Smith et al. 2001; Foster et al. 2012). We used an augmented version of the atomic data to include inner-shell processes and updated Fe-L lines (see Badenes et al. 2006).  For the Shell region we extracted a background spectrum from a source free area outside the southern boundary of the SNR.  To fit the spectrum of the Shell region we varied the abundances for O, Ne, Mg, Si, and Fe.  We fixed the S abundance at the solar value \citep{ande89} because there are few counts above ${\it E}$ $\sim 2$ keV.  This fit is statistically acceptable ($\chi^{2}_{\nu}$ $\sim$ 1) with abundance values ranging from 0.3--0.5 (abundances are with respect to solar hereafter).  We estimate the electron temperature to be $kT$ $\sim$ 0.5 keV with an ionization timescale of $n_et$ $\sim$ 1.6 $\times$ 10$^{11}$ cm$^{-3}$ s.  Our best-fit abundance values are consistent with those measured by Park et al. (2007) for the outer shell regions.  We also fit the Shell spectrum with all abundance values fixed at solar.  This fit is statistically acceptable although the reduced $\chi^{2}$ is somewhat higher ($\chi^{2}_{\nu}$ $\sim$ 1.3) than that for our previous fit.  In either fit the swept-up CSM/ISM (with no metal overabundances) is implied for the Shell region.  Hereafter we use our best-fit model abundances for the Shell region (Table 1) because the statistical improvement is significant (F-probability $\sim$ 2.5 $\times$ 10$^{-5}$).          

We initially fit the North, South, and West regions with a single plane shock model ($\chi^{2}_{\nu}$ $\sim$ 1.3, 1.1, and 1.2 for the North, South, and West regions, respectively).  We selected background from a source free region outside of the SNR.  We note that, the background spectrum does not take into account emission originating from the ISM/CSM components from the outer shells of the SNR along the line of sight.  Thus, we added a second NEI shock component to our model (Fig. 3) to consider the superposed outer shell spectrum.  All model parameters (except for the volume emission measure [EM]) for the outer shell component were fixed at the best-fit values that we obtained from the Shell region.  Our two-component shock model fits are equally acceptable ($\chi^{2}_{\nu}$ $\sim$1.25, 1.12, and 1.16 for the North, South and West regions, respectively) as those with the one-component shock model fits.  The electron temperatures and ionization timescales, as well as the best-fit abundances for Ne, Si, Fe, and S in the ejecta component are consistent (within uncertainties) with those estimated from the our single-shock model fits.  The contribution from the superimposed shell is $\sim$16--20\% of the total flux for these regions.  In our two-component model fits the best-fit O abundance is negligible in the North and South regions while it remains similar in the West region to that estimated with our single-shock model fit.  Although the two-shock model fits are not statistically distinguishable from the one-shock model fits based on F-test (F-probability $\sim$0.1, 0.2, and 0.3 for the North, South and West regions, respectively), they represent a physically more realistic model.  Hereafter, we discuss the ejecta nature based on the results from our two-component shock model fits (Table 1).         

\section{\label{sec:disc} DISCUSSION \& CONCLUSIONS}

In Fig. 4 we compare the average abundance ratios from our North, South and West regions to the predicted nucleosynthesis yields of various Type Ia \citep{iwam99} and core-collapse (CC) SN models (for progenitor masses of 13-40 $M_{\sun}$, Nomoto et al. 2006).  The observed abundances are more consistent with Type Ia models rather than CC models.  This is largely due to the lack of O-group elements (coupled with enhanced Fe) in the detected metal-rich ejecta.  This ejecta composition is characteristic for Type Ia SN and rules out a massive progenitor for G299.2-2.9, confirming the conclusions by Park et al. (2007).  Our measured Fe/Si abundance ratio is in plausible agreement with delayed-detonation models for Type Ia SNe, while the O/Si and Ne/Si abundance ratios appear to be in agreement with either delayed-detonation or deflagration models.  The S/Si ratio is not effective in discriminating between Type Ia models.  Although we place only upper limits on the O and Ne abundances, and thus cannot provide tight constraints on the explosion physics, our estimated ejecta abundance ratios generally suggest that G299.2-2.9 was the remnant of a Type Ia SN explosion through a delayed-detonation process. 

A significant elongation of Si- and Fe-rich ejecta continuously extends from the center of the SNR out to the western outermost boundary.  The central ejecta show a differential composition between the northern and southern halves, possibly indicating different layers of the Si-burning \citep{thie86}.   These non-uniform substructures of ejecta are unlike those found in other Type Ia SNRs with a similar age to G299.2-2.9.  For instance DEM L71, a Type Ia SNR in the Large Magellanic Cloud, shows a nearly circular (or somewhat elliptical) central ejecta emission feature \citep{hugh03}.  DEM L71's central ejecta shows a radial stratification between Si and Fe \citep{hugh03} which is expected from standard Type Ia SN models.  The Galactic Type Ia SNR G337.2-0.7 shows a complex X-ray morphology with faint emission features (probably metal-rich ejecta) surrounding the central ejecta region \citep{rako06}. Although photon statistics are limited in the data for G337.2-0.7, significant spatial variations of ejecta elements were not observed there \citep{rako06}.  Although in a different stage of dynamical evolution, young, well-observed ``canonical'' Type Ia SNRs like Tycho or SN1006 do not exhibit similar ejecta structures as seen in G299.2-2.9\footnote{Mild asymmetries in the ejecta have been observed in SN 1006 by Uchida et al. (2013) and Winkler et al. (2014).}.  (The two large knots, one Fe-rich and the other Si-rich, along the southeast rim of Tycho \citep{vanc95} might possibly suggest a dynamically younger stage of the ejecta elongation we observe in G299.2-2.9, but it is only speculative.)          

These structured metal-rich ejecta features in G299.2-2.9 might have been caused by an asymmetric Type Ia SN.  Asymmetric Type Ia explosions have been suggested by a growing number of SD models in which detonations may ignite at multiple off-center positions in the progenitor (e.g., Gamezo et al, 2005; Maeda et al. 2010; Malone et al. 2014).  Asymmetries are also suggested to occur in the double-detonation of SD sub-Chandrasekhar SNe \citep{fink10}.  Collisional DD scenarios also predict asymmetric explosions \citep{kush13}.  On the other hand spectropolarimetric observations of Type Ia SNe indicate that most show negligible continuum polarization, suggesting that typical Type Ia SNe are spherically symmetric (see Maoz et al. 2013).  It has also been shown that Type Ia SNRs exhibit stronger spherical and mirror symmetries based upon their X-ray morphology than core-collapse SNRs (Lopez et al. 2009,2011).  However, some Type Ia SNe show strong line polarization, particularly for the [Ca II] and [Si II] lines, before reaching the maximum light, suggesting that portions of the ejecta are asymmetrically distributed (Wang \& Wheeler 2008; Maund et al. 2010).  Foley et al. (2012) found that SNe Ia with blue-shifted narrow Na D profiles (an indicator of circumstellar material interaction) tend to have higher ejecta velocities than those with no Na D absorption or those with red-shifted single or symmetric profiles.  One possible explanation for these higher ejecta velocities is an asymmetric explosion \citep{fole12}.  Thus, our observed asymmetrical ejecta distribution in G299.2-2.9, particularly its elongation to the western outermost boundary, might have originated from a significantly asymmetric Type Ia explosion.            
  
Alternatively, the observed asymmetric ejecta may be a result of the explosion in a non-uniform ISM/CSM.  A CSM-ejecta interaction can be considered in the context of a SD scenario in which the ambient medium has been modified by stellar winds from the companion or progenitor, similar to that of Kepler's SNR (e.g, Chiotellis et al. 2012, Patnaude et al. 2012, Burkey et al. 2013) or RCW 86 \citep{broe14}, but see Soker (2014) for a discussion of DD CSM scenarios.  A non-uniform environment surrounding G299.2-2.9 \citep{park07} would suggest that this SNR may have been a remnant of the Type Ia-CSM class such as SNe 2002ic, 2005gj, 2008J, and PTF 11kx (Hamuy et al. 2003, Aldering et al. 2006, Soker et al. 2013), in which Type Ia SNe are interacting with dense CSM.  Observations show that $\gtrsim20\%$ of Type Ia SNe (including Type Ia-CSM) may be interacting with CSM released by the progenitor system prior to the explosion (Sternberg et al. 2011, Foley et al. 2012, Maguire et al. 2013).  However, a majority of Type Ia SNe appear to expand into {\it uniform} surroundings (Badenes et al 2007, Yamaguchi et al. 2014), and G299.2-2.9 may represent a relatively small population of Type Ia SNe that interact with modified CSM.  We note that planetary nebula-like bi-polar outflows from the progenitor or companion (suggested by Tsebrenko \& Soker 2013) are unlikely the cause of the observed ejecta elongation in G299.2-2.9, because such outflows would channel ejecta through a bi-polar stream as well, while G299.2-2.9 shows only a {\it one-sided} extension primarily toward the west.  Our measured density for the western outermost boundary is similar to those for the faint outermost emission features surrounding the SNR, which is significantly lower than that of the bright inner shells \citep{park07}.  Thus the observed ejecta elongation toward the west might be due to this density gradient. However, significant ejecta emission similarly extending toward the faint outermost boundary in non-western directions is not clearly observed.  Thus, it is unclear whether the observed western elongation of the ejecta was caused entirely by this density gradient.  If a lower ambient density was the main cause of the elongation to the west, the origin of such a low density there remains uncertain (e.g., the progenitor system, local interstellar structure, etc).  A detailed spatially-resolved study of the ejecta and ISM/CSM features together with 3-D hydrodynamic simulations of asymmetric Type Ia explosions in various non-uniform environments are required to reveal the true origin of this peculiar Type Ia SNR.

\acknowledgments

This work has been supported in part by NASA under the {\it Chandra} grant GO0-11076X and NASA Contract NAS8-03060.

\clearpage

\begin{figure}
\figurenum{1}
\centerline{\includegraphics[angle=0,width=\textwidth]{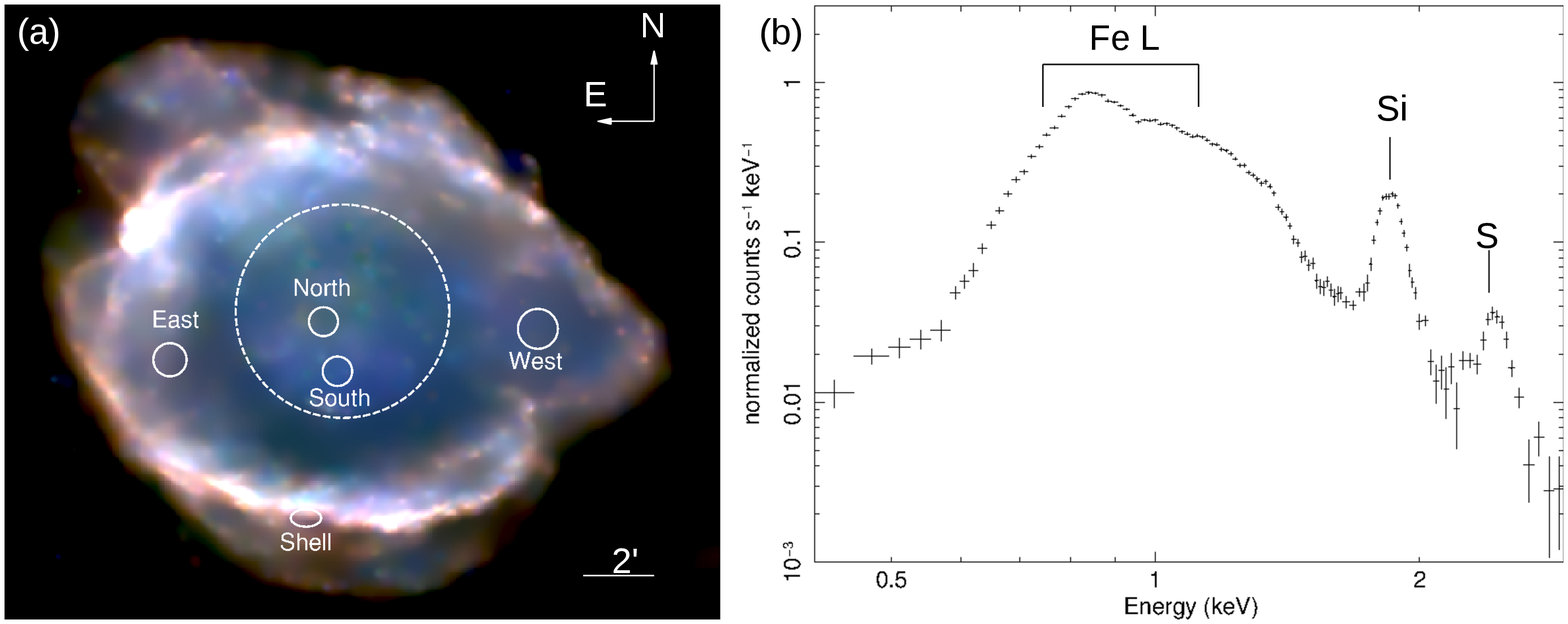}}
\figcaption[]{(a) An exposure-corrected 3-color image of G299.2-2.9 based on our {\it Chandra} data. Red, green, and blue represent the 0.4--0.72, 0.72--1.4, and 1.4--3.0 keV bands, respectively.  Each sub-band image has been binned by 4 $\times$ 4 pixels and adaptively smoothed for the purposes of display. Regions in which we extract characteristic spectra are marked. (b) The overall spectrum of the central ejecta nebula as extracted from the dashed circle in (a).  Several atomic emission line features are marked.     
\label{fig:fig1}}
\end{figure}

\begin{figure}
\figurenum{2}
\centerline{\includegraphics[angle=0,width=\textwidth]{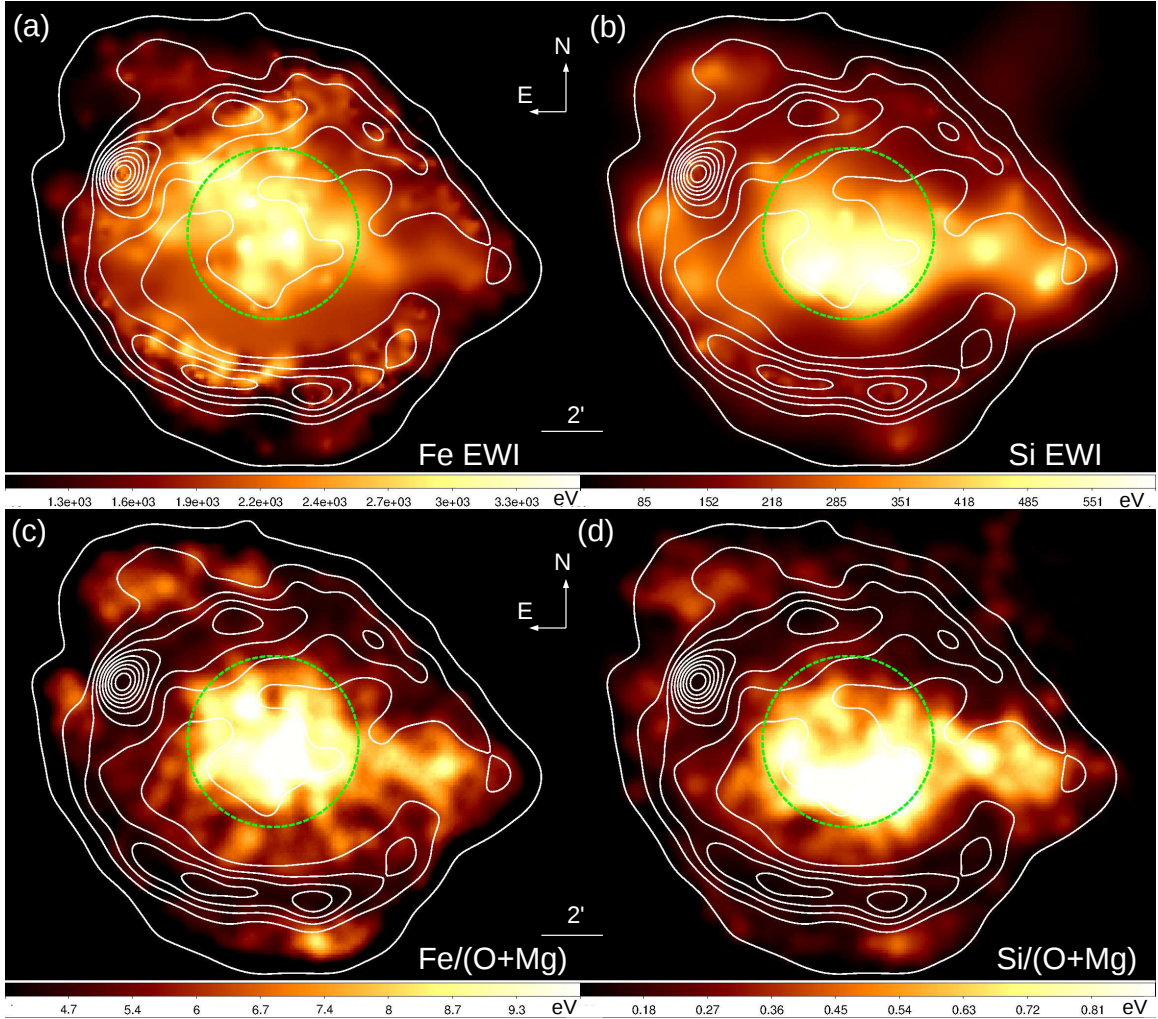}}
\figcaption[]{(a) Fe-L line EW image of G299.2-2.9. We extracted line emission from the 0.75--1.15 keV band with a low continuum from the 0.5--0.6 keV and the high continuum from the 1.20--1.28 keV bands, respectively.  (b) Si line EW image of G299.2-2.9.  We extracted line emission from the 1.78--1.93 keV band with a low continuum from the 1.5--1.68 keV and a high continuum from the 2--2.3 keV bands, respectively.  (c) Fe (0.75--1.15 keV) to O+Mg (0.62--0.7 keV + 1.28--1.38 keV) line ratio map.  (D) Si (1.78--1.93 keV) to O+Mg line ratio map.  We binned all images by 8 $\times$ 8 pixels and then adaptively smoothed. In (a)-(d), images are overlaid with contours from the broadband image (0.4--3.0 keV) of the SNR.
\label{fig:fig2}}
\end{figure}

\begin{figure}
\figurenum{3}
\centerline{\includegraphics[angle=0,width=\textwidth]{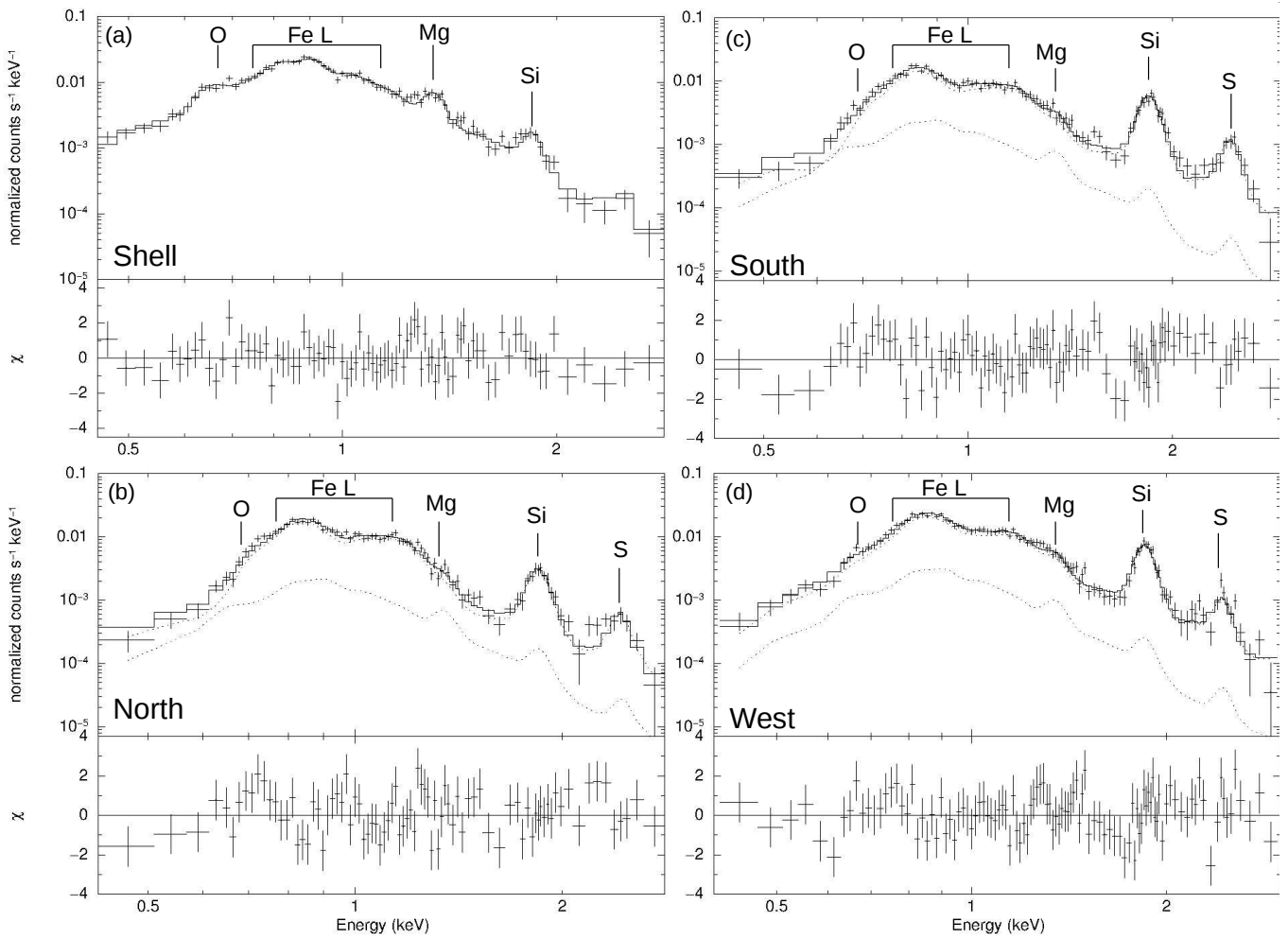}}
\figcaption[]{ ACIS spectra from characteristic regions of G299.2-2.9. (a) The Shell region.  (b) The North region. (c) The South region. (d)  The West region.  The best-fit two-component shock model is overlaid in each panel. In (a)-(d), residuals from the best-fit model are plotted in the bottom panel. 
\label{fig:fig3}}
\end{figure}

\begin{deluxetable}{ccccc}
\footnotesize
\tablecaption{Best-fit parameters for characteristic regions in G299.2-2.9.
\label{tbl:tab1}}
\tablewidth{0pt}
\tablehead{\colhead{Parameters} & \colhead{North} & \colhead{South} & \colhead{West} & \colhead{Shell}} 
\startdata
$N_{H}$ ($\times$ 10$^{22}$ cm$^{-2}$) & 0.30$^{+0.03}_{-0.04}$ & 0.31$^{+0.03}_{-0.04}$ & 0.37$^{+0.05}_{-0.05}$ & 0.26$^{+0.12}_{-0.06}$ \\
$kT$ (keV) & 1.36$^{+0.02}_{-0.07}$ & 1.36$^{+0.02}_{-0.15}$ & 1.31$^{+0.08}_{-0.16}$ & 0.54$^{+0.08}_{-0.17}$\\
O & $<$ 0.21\tablenotemark{a} & $<$ 0.20\tablenotemark{a} & 0.34$^{+0.31}_{-0.19}$ & 0.32$^{+0.11}_{-0.09}$ \\
Ne & $<$ 0.72\tablenotemark{a} & $<$ 0.12\tablenotemark{a} & $<$ 0.09\tablenotemark{a} & 0.43$^{+0.10}_{-0.08}$ \\ 
Si & 5.77$^{+6.17}_{-3.01}$ & 7.53$^{+3.93}_{-2.06}$ & 4.17$^{+2.30}_{-1.14}$ & 0.49$^{+0.16}_{-0.14}$ \\
S & 15.80$^{+18.47}_{-4.25}$ & 18.18$^{+11.15}_{-7.42}$ & 5.53$^{+3.92}_{-2.06}$ & 1\tablenotemark{b} \\
Fe & 6.21$^{+11.93}_{-1.82}$ & 3.73$^{+1.18}_{-1.23}$ & 2.36$^{+0.92}_{-0.46}$ & 0.47$^{+0.14}_{-0.15}$ \\ 
{\it $n_et$} ($\times$ 10$^{10}$ cm$^{-3}$ s) & 1.69$^{+0.41}_{-0.63}$ & 2.09$^{+0.58}_{-0.25}$ & 3.13$^{+0.87}_{-0.53}$ &  15.90$^{+12.40}_{-5.20}$\\ 
EM$_{1}$\tablenotemark{c} ($\times$ 10$^{54}$ cm$^{-3}$)  & 2.57$^{+0.66}_{-1.42}$  & 3.28$^{+1.92}_{-1.06}$ &  6.96$^{+3.09}_{-2.56}$ & --  \\
EM$_{2}$\tablenotemark{d} ($\times$ 10$^{54}$ cm$^{-3}$) & 3.79$^{+3.20}_{-4.05}$  & 4.37$^{+3.33}_{-2.61}$ & 6.11$^{+4.35}_{-3.77}$ &  3.68$^{+6.56}_{-1.53}$\\
$\chi^{2}$/dof & 93.49/75 & 96.47/86 & 114.44/99 & 78.97/78\\
\enddata
\tablecomments{Uncertainties are at 90\% confidence level.}
\tablenotetext{a}{90\% upper limit.}
\tablenotetext{b}{Fixed at solar abundance}
\tablenotetext{c}{Ejecta-component EM}
\tablenotetext{d}{ISM/CSM-component EM}
\end{deluxetable}

\begin{figure}
\figurenum{4}
\centerline{\includegraphics[angle=0,width=\textwidth]{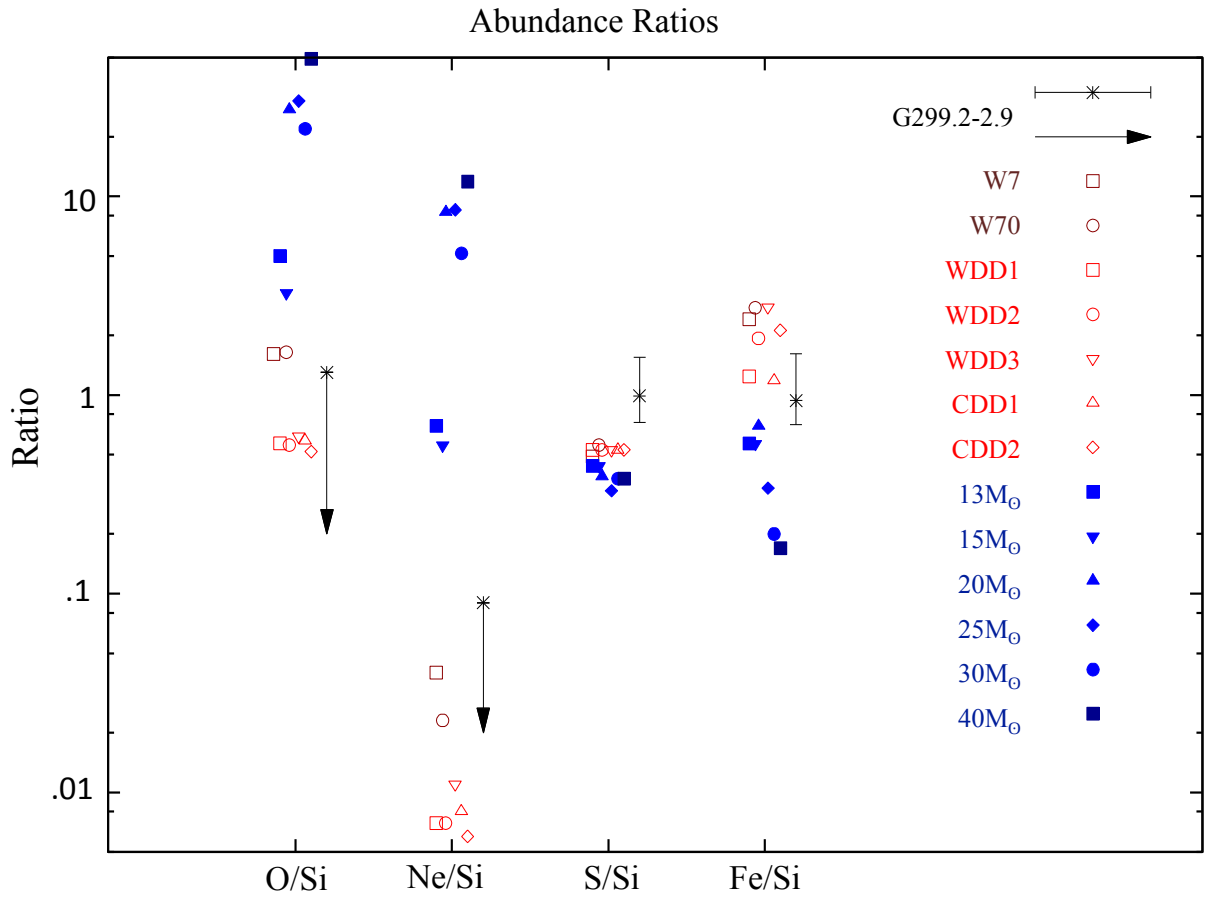}}
\figcaption[]{Abundance ratios of Type Ia supernovae (Iwamoto et al. 1999; red and brown markers).  For comparison, we include core-collapse SN models (blue markers) for the progenitor mass range 13-40 M$_{\sun}$ (Nomoto et al. 2006).   
\label{fig:fig4}}
\end{figure}


\begin{thebibliography}{}

\bibitem[Aldering et al. 2006]{alde06} Aldering, G., et al., 2006, \apj, 650, 510

\bibitem[Anders \& Grevesse 1989]{ande89} Anders, E., \& Grevesse, N., 1989, Geochimica et Cosmochimica Acta, 53, 197

\bibitem[Badenes et al. 2007]{bade07} Badenes, C., Hughes, J., Bravo, E., \& Langer N., 2007, \apj, 662, 472 

\bibitem[Borkowski et al. 2001]{bork01} Borkowski, K., Lyerly, W., \& Reynolds, S. 2001, \apj, 548, 820

\bibitem[Broersen et al. 2014]{broe14} Broersen, S., Chiotellis, A., Vink, J., \& Bamba, A., 2014, MNRAS, in press (arXiv:1404.5434)  

\bibitem[Burkey et al. 2013]{burk13} Burkey, M., Reynolds, S., Borkowski, K., \& Blondin, J., 2013, \apj, 764, 63

\bibitem[Busser \& Aschenbach 1995]{buss95} Busser, J.-U., \& Aschenbach, B. 1995, IAU Circ. 6142

\bibitem[Chiotellis et al. 2012]{chio12} Chiotellis, A., Schure, K., \& Vink, J., 2012, A\&A, 537, 139

\bibitem[Dilday et al. 2012]{dild12} Dilday, B., et al., 2012, Science, 337, 942

\bibitem[Fink et al. 2010]{fink10} Fink, M., R\"opke, F., Hillebrandt, W., Seitenzahl, I., Sim, S., \& Kromer, M., 2010, A\&A, 514, 53

\bibitem[Foley et al. 2012]{fole12} Foley, R., et al., 2012, \apj, 752, 101

\bibitem[Foster et al. 2012]{fost12} Foster A., Ji, L., Smith, R., \& Brickhouse, N., 2012, \apj, 756, 128

\bibitem[Gamezo et al. 2005]{game05} Gamezo, V., Khokhlov, A., \& Oran, E., 2005, \apj, 623, 337

\bibitem[Garmire et al. 2003]{garm03} Garmire, G., Bautz, M., Ford, P., Nousek, J., \& Ricker, Jr., G. 2003, in “X-Ray and Gamma-Ray Telescopes and Instruments for Astronomy”, Proc. of SPIE, eds. J. E. Tr\"umper and H. D. Tananbaum, 4851, 28

\bibitem[Hamuy et al. 2003]{hamu03} Hamuy, M., 2003, Nature, 424

\bibitem[Hughes et al. 2003]{hugh03} Hughes, J., Ghavamian, P., Rakowski, C., \& Slane, P., 2003, \apj, 582, L95

\bibitem[Hughes et al. 2007]{hugh07} Hughes, J., Chugai, N., Chevalier, R., Lundqvist, P., \& Schlegel, E., 2007, \apj, 670, 1260 

\bibitem[Hwang et al. 2000]{hwan00} Hwang, U., Holt, S., \& Petre, R., 2000, \apj, 537, 119

\bibitem[Iwamoto et al. 1999]{iwam99} Iwamoto, K., Brachwitz, F., Nomoto, K., Kishimoto, N., Umeda, H., Hiz, R., \& Thielemann, F., 1999, \apj, 125, 439

\bibitem[Kushnir et al. 2013]{kush13} Kushnir, D., Katz, B., Dong, S., Livne, E., \& Fernandez, R., 2013 \apj, 778, L37

\bibitem[Lopez et al. 2009]{lope09} Lopez, L., Ramirez-Ruiz, E., Badenes, C., Huppenkothen, D., Jeltema, T., \& Pooley, D., 2009, \apj, 706, L106

\bibitem[Lopez et al. 2011]{lope11} Lopez, L., Ramirez-Ruiz, E., Badenes, C., Huppenkothen, D., \& Pooley, D., 2011, \apj, 732, 114 

\bibitem[Maeda et al. 2010]{maed10} Maeda, K., et al., 2010, Nature, 466, 82

\bibitem[Maguire et al 2013]{magu13} Maguire, K., et al., 2013, MMRAS, 436, 222

\bibitem[Malone et al. 2014]{malo14} Malone, C., Nonaka, A., Woosley, S., Almgren, A., Bell, J., Dong, S., \& Zingale, M., 2013, \apj, 782, 11 

\bibitem[Maoz et al. 2014]{maoz14} Maoz, D., Mannucci, F., \& Nelemans, G., 2014, ARA\&A, Submitted (arXiv:1312.0628)  

\bibitem[Maund et al. 2010]{maun10} Maund, J., H\"oflich, P., Patat, F., Wheeler, C., Zelaya, P., Baade, D., Wang, L., Clocchiatti, A., \& Quinn, J., 2010, \apj, 725, L167

\bibitem[Nomoto et al. 2006]{Nomo06} Nomoto, K., Tominaga, N., Umeda, H., Kobayashi, C., \& Maeda, K., 2006, Nuclear Physics A, 777, 424

\bibitem[Park et al. 2002]{park02} Park, S., Roming, P., Hughes, J., Slane, P., Burrows, D., Garmire, G., \& Nousek, J. 2002, \apj, 564, L39

\bibitem[Park et al. 2007]{park07} Park, S., Slane, P., Hughes, J., Mori, K., Burrows, D., \& Garmire, G. P. 2007, \apj, 665, 1173

\bibitem[Patnaude et al. 2012]{patn12} Patnaude, D., Badenes, C., Park, S., \& Laming, J., 2012, \apj, 756, 6

\bibitem[Rakowski et al. 2006]{rako06} Rakowski, C., Badenes, C., Gaensler, B., Gelfand, J., Hughes, J., \& Slane, P., 2006, \apj, 646, 982 

\bibitem[Slane et al. 1996]{slan96} Slane, P., Vancura, O., \& Hughes, J., 1996, \apj, 465, 840

\bibitem[Smith et al. 2001]{smith01} Smith, R., Brickhouse, N., Liedahl, D., \& Raymond, J. 2001, \apj, 556, L91
 
\bibitem[Soker et al. 2013]{soke13} Soker, N., Kashi, A., Garcia-Berro, E., Torres, S., \& Camacho, J., 2013, MNRAS, 430, 1970

\bibitem[Soker et al. 2014]{soke14} Soker, N., 2014 Submitted (arXiv:1405.0173)

\bibitem[Sternberg et al. 2011]{ster11} Sternberg, A., et al., 2011, Science, 333, 856

\bibitem[Thielemann et al. 1986]{thie86} Thielemann, F., Nomoto, K., \& Yokoi, K., 1986, A\&A, 158, 17

\bibitem[Townsley et al. 2000]{town00} Townsley, L., Broos, P., Garmire, G., \& Nousek, J. 2000, \apj, 534, L139

\bibitem[Tsebrenko \& Soker 2013]{tseb13} Tsebrenko, D., \& Soker, N. 2013, MNRAS, 435, 320

\bibitem[Uchida et al. 2013]{uchi13} Uchida, H., Yamaguchi, H., \& Koyama, K., 2013, \apj, 771, 56 

\bibitem[Vancura et al. 1995]{vanc95} Vancura, O., Gorenstein, P., \& Hughes, J,. 1995, \apj, 441, 680

\bibitem[Wang \& Wheeler 2008]{wang08} Wang, L., \& Wheeler, J., 2008, ARA\&A 46, 433

\bibitem[Winkler et al. 2014]{wink14} Winkler, F., Williams, B., Reynolds, S., Petre, R., Long, K., Katsuda, S., \& Hwang, U., 2014, \apj, 781, 65

\bibitem[Yamaguchi et al. 2014]{yama14} Yamaguchi et al., 2014, \apj, 785L, 27 

\end{thebibliography}
\end{document}